\documentclass{article}
\usepackage{amsmath}

\input{tcilatex}

\begin{document}

\title{Discussing the Relationship between the Static and Dynamic Light Scattering }
\author{Yong Sun}
\maketitle

\begin{abstract}
Both the static $\left( SLS\right) $ and dynamic $\left( DLS\right) $ light
scattering techniques are used to obtain the size information from the
scattered intensity, but the static radius $R_{s}$ and the apparent
hydrodynamic radius $R_{h,app}$ are different. In this paper, the
relationship between SLS and DLS is discussed using dilute water dispersions
of two different homogenous spherical particles, polystyrene latexes and
poly($N$-isopropylacrylamide) microgels, with a simple assumption that the
hydrodynamic radius $R_{h}$ is in proportion to the static radius $R_{s}$,
when Rayleigh-Gans-Debye approximation is valid. With the assistance of the
simulated data, the apparent hydrodynamic radius $R_{h,app}$ has been
discussed. The results show that the apparent hydrodynamic radius is
different with the mean hydrodynamic radius of particles and is a composite
size obtained from averaging the term $\exp \left( -q^{2}D\tau \right) $ in
the static size distribution $G\left( R_{s}\right) $ with the weight $%
R_{s}^{6}P\left( q,R_{s}\right) $.
\end{abstract}

It is known that the structural information and mass weight are included in
the relationship between the average scattered intensity and the scattering
angle. Measuring this dependence is called the static light scattering $%
\left( SLS\right) $. The analysis of the time auto-correlation of the
scattered light intensity can provide the dynamic information of particles,
called the dynamic light scattering $\left( DLS\right) $. SLS obtains the
information from the optical features and DLS gets the information from both
the optical and hydrodynamic characteristics of particles. If the
relationship between the optical and hydrodynamic quantities of particles
can be built, the experimental values of the normalized time
auto-correlation function of the scattered light intensity $g^{\left(
2\right) }\left( \tau \right) $ can be expected using the size information
obtained from SLS. In this article, the relationship between the SLS and DLS
for homogenous spherical particles is discussed with a simple assumption
when Rayleigh-Gans-Debye $\left( RGD\right) $ approximation is valid.

For homogeneous spherical particles where the RGD approximation is valid,
the normalized time auto-correlation function of the electric field of the
scattered light $g^{\left( 1\right) }\left( \tau \right) $ is given by

\begin{equation}
g^{\left( 1\right) }\left( \tau \right) =\frac{\int_{0}^{\infty
}R_{s}^{6}P\left( q,R_{s}\right) G\left( R_{s}\right) \exp \left(
-q^{2}D\tau \right) dR_{s}}{\int_{0}^{\infty }R_{s}^{6}P\left(
q,R_{s}\right) G\left( R_{s}\right) dR_{s}},  \label{Grhrs}
\end{equation}
where $q$ is the scattering vector, $R_{s}$ is the static radius, $\tau $ is
the delay time, $D$ is the diffusion coefficient, $G\left( R_{s}\right) $ is
the number distribution and the form factor $P\left( q,R_{s}\right) $ is

\begin{equation}
P\left( q,R_{s}\right) =\frac{9}{q^{6}R_{s}^{6}}\left( \sin \left(
qR_{s}\right) -qR_{s}\cos \left( qR_{s}\right) \right) ^{2}.  \label{factor}
\end{equation}
In this discussion, the number distribution is chosen as a Gaussian
distribution

\begin{equation*}
G\left( R_{s};\left\langle R_{s}\right\rangle ,\sigma \right) =\frac{1}{%
\sigma \sqrt{2\pi }}\exp \left( -\frac{1}{2}\left( \frac{R_{s}-\left\langle
R_{s}\right\rangle }{\sigma }\right) ^{2}\right) ,
\end{equation*}
where $\left\langle R_{s}\right\rangle $ is the mean static radius and $%
\sigma $ is the standard deviation relative to the mean static radius.

From the Stokes-Einstein relation

\begin{equation*}
D=\frac{k_{B}T}{6\pi \eta _{0}R_{h}},
\end{equation*}
where $\eta _{0}$, $k_{B}$, $T$ and $R_{h}$ are the viscosity of the
solvent, Boltzmann's constant, absolute temperature and hydrodynamic radius
of a particle, respectively, the hydrodynamic radius can be obtained.

For simplicity, we assume that the relationship between the static and
hydrodynamic radii is given by 
\begin{equation}
R_{h}=aR_{s},  \label{RsRh}
\end{equation}
where $a$ is a constant. With the function between the normalized time
auto-correlation function of the scattered light intensity $g^{\left(
2\right) }\left( \tau \right) $ and the normalized time auto-correlation
function of the electric field of the scattered light $g^{\left( 1\right)
}\left( \tau \right) $ $\left[ 1\right] $

\begin{equation}
g^{\left( 2\right) }\left( \tau \right) =1+\beta \left( g^{\left( 1\right)
}\right) ^{2},  \label{G1G2}
\end{equation}
the relationship between SLS\ and DLS is built and the values of the
normalized time auto-correlation function of the scattered light intensity $%
g^{\left( 2\right) }\left( \tau \right) $ can be expected.

In this paper, the calculated and experimental values of $g^{\left( 2\right)
}\left( \tau \right) $ for two samples were compared. One is the polystyrene
latex sample\ with the normalized size information: the mean radius is 33.5
nm and the standard deviation is 2.5 nm provided by the supplier, from
Interfacial Dynamics Corporation (Portland, Oregon).\ The sample was diluted
for light scattering to weight factor of $1.02\times 10^{-5}$ with fresh
de-ionized water from a Milli-Q plus water system (Millipore, Bedford, with
a 0.2 $\mu m$ filter). The other is the poly($N$-isopropylacrylamide)
(PNIPAM) microgel sample with the molar ratio 1\% of crosslinker $%
N,N^{^{\prime }}$-methylenebisacrylamide over $N$-isopropylacrylamide. The
PNIPAM microgel sample was diluted to $8.56\times 10^{-6}$.\ The size
information was obtained fitting the SLS data. At a temperature of 302.33 K,
the mean static radius is 254.3$\pm $0.1nm, the standard deviation is 21.5$%
\pm $0.3nm and $\chi ^{2}$ is 2.15 $\left[ 2\right] .$

If the constant $a$ for the polystyrene latex sample is assumed to be 1.1
and the size information provided by the supplier is thought to be
consistent with that obtained from SLS, all the experimental and calculated
values of $g^{\left( 2\right) }\left( \tau \right) $ at a temperature of
298.45 K and the scattering angles 30$^{\text{o}}$, 60$^{\text{o}}$, 90$^{%
\text{o}}$, 120$^{\text{o}}$ and 150$^{\text{o}}$ are shown in Fig. 1.a.
When the constant $a$ for the PNIPAM sample is assumed to be 1.21, all the
experimental and calculated values at the scattering angles 30$^{\text{o}}$,
50$^{\text{o}}$ and 70$^{\text{o}}$ are shown in Fig. 1.b. Figure 1 shows
that the calculated values are consistent with the experimental data very
well.

If the expected values were calculated using Bargeron's equation [3], all
the experimental and calculated values of $g^{\left( 2\right) }\left( \tau
\right) $ for the polystyrene latex sample at the scattering angles 30$^{%
\text{o}}$, 60$^{\text{o}}$, 90$^{\text{o}}$, 120$^{\text{o}}$ and 150$^{%
\text{o}}$ are shown in Fig. 2.a; all the experimental and calculated values
for the PNIPAM sample at the scattering angles 30$^{\text{o}}$, 50$^{\text{o}%
}$ and 70$^{\text{o}}$ are shown in Fig. 2.b. Figure 2 shows that the
expected values have large differences with the experimental data.

Traditionally the size information is obtained\ from DLS. The standard
method is the cumulant or the inverse Laplace transform. For the five
experimental data of $g^{\left( 2\right) }\left( \tau \right) $ measured
under the same conditions as the SLS data, their corresponding fit results
of $g^{\left( 2\right) }\left( \tau \right) $ using the first cumulant and
first two cumulant $\left[ 4,5\right] $ respectively for the PNIPAM microgel
sample at a temperature of 302.33 K and a scattering angle of 30$^{\text{o}}$
are listed in Table 1.

\begin{center}
$\underset{\text{Table.1 The fit results for the PNIPAM sample at a
temperature of 302.33 K and a scattering angle of 30}^{\text{o}}.}{
\begin{tabular}{|c|c|c|c|c|c|}
\hline
& $\left\langle \Gamma \right\rangle _{first}$ & $\chi ^{2}$ & $\left\langle
\Gamma \right\rangle _{two}$ & $\mu _{2}$ & $\chi ^{2}$ \\ \hline
1 & 79.5$\pm $0.1 & 0.07 & 79.9$\pm $0.3 & 28.20$\pm $15.99 & 0.04 \\ \hline
2 & 79.0$\pm $0.1 & 0.33 & 80.4$\pm $0.3 & 90.10$\pm $17.11 & 0.04 \\ \hline
3 & 79.7$\pm $0.1 & 0.11 & 80.3$\pm $0.3 & 39.17$\pm $16.19 & 0.05 \\ \hline
4 & 79.4$\pm $0.1 & 0.07 & 79.7$\pm $0.3 & 20.93$\pm $15.92 & 0.06 \\ \hline
5 & 78.7$\pm $0.1 & 0.53 & 80.4$\pm $0.3 & 112.75$\pm $17.26 & 0.08 \\ \hline
\end{tabular}
\ }$
\end{center}

From the fit results, the values of the mean decay constant $\left\langle
\Gamma \right\rangle $ show an independence on the measurements, but the
results of $\mu _{2}$ have a strong dependence on the measurements. The
values of $\mu _{2}$ are often negative. It's a contradiction with its
definition. In order to discuss this problem conveniently, the simulated
data are used.

The simulated data were produced using the size information: the mean static
radius is 260 nm and the standard deviation is 26 nm. the temperature $T$
was set to 302.33K, the viscosity $\eta _{0}$ of the solvent was 0.8132 mPa$%
\cdot $S, the scattering angle was 30$^{\text{o}}$ and the constant $a$ was
chosen as 1.2. When the data of $\left( g^{\left( 2\right) }\left( \tau
\right) -1\right) /\beta $ were obtained, the 1\% statistical noises were
added and the random errors were set 3\%. Five simulated data were produced
respectively. The fit results for the five simulated data using\ the first
cumulant and first two cumulant respectively are shown in Table 2.

\begin{center}
$\underset{\text{Table 2 The fit results for the simulated data with the
standard deviation 26 nm.}}{
\begin{tabular}{|c|c|c|c|c|c|}
\hline
& $\left\langle \Gamma \right\rangle _{first}$ & $\chi ^{2}$ & $\left\langle
\Gamma \right\rangle _{two}$ & $\mu _{2}$ & $\chi ^{2}$ \\ \hline
1 & 79.27$\pm $0.01 & 11.98 & 79.90$\pm $0.02 & 22.0$\pm $0.6 & 7.75 \\ 
\hline
2 & 78.50$\pm $0.01 & 4.56 & 78.98$\pm $0.03 & 9.8$\pm $0.6 & 3.83 \\ \hline
3 & 78.35$\pm $0.01 & 5.67 & 79.43$\pm $0.06 & 20.6$\pm $1.2 & 4.66 \\ \hline
4 & 78.33$\pm $0.01 & 25.40 & 78.25$\pm $0.02 & -1.7$\pm $0.4 & 25.44 \\ 
\hline
5 & 78.596$\pm $0.004 & 15.75 & 78.79$\pm $0.02 & 4.9$\pm $0.5 & 15.55 \\ 
\hline
\end{tabular}
\ }$
\end{center}

From the fit results\ of the simulated data that are shown in Table 2, the
situation is the same as the experimental data, the values of the mean decay
constant $\left\langle \Gamma \right\rangle $ show an independence on the
different noises and errors, and the results of $\mu _{2}$ have a strong
dependence on them. The values of $\mu _{2}$ can be negative. As we have
discussed, a truncated Gaussian distribution can give better results for the
SLS data of the PNIPAM sample at a temperature of 302.33 K $\left[ 2\right] $%
, so the five simulated data were produced respectively again with the
truncated Gaussian distribution that the range of integral is 221 to 299 nm.
The fit results for this five simulated data using\ the first cumulant and
first two cumulant respectively are shown in Table 3. The values of the
quantity $\mu _{2}$ still have large differences for different simulated
data and are often negative.

\begin{center}
$\underset{\text{Table 3 The fit results for the simulated data with the
truncated distribution.}}{
\begin{tabular}{|c|c|c|c|c|c|}
\hline
& $\left\langle \Gamma \right\rangle _{first}$ & $\chi ^{2}$ & $\left\langle
\Gamma \right\rangle _{two}$ & $\mu _{2}$ & $\chi ^{2}$ \\ \hline
1 & 79.996$\pm $0.002 & 10.96 & 79.73$\pm $0.02 & -5.4$\pm $0.4 & 10.36 \\ 
\hline
2 & 79.83$\pm $0.01 & 20.57 & 79.79$\pm $0.05 & -0.95$\pm $1.24 & 20.63 \\ 
\hline
3 & 80.091$\pm $0.004 & 5.30 & 80.69$\pm $0.04 & 12.8$\pm $0.8 & 4.61 \\ 
\hline
4 & 79.926$\pm $0.009 & 3.97 & 80.14$\pm $0.03 & 4.3$\pm $0.6 & 3.84 \\ 
\hline
5 & 79.985$\pm $0.005 & 9.00 & 80.48$\pm $0.02 & 9.4$\pm $0.3 & 5.51 \\ 
\hline
\end{tabular}
\ }$
\end{center}

Comparing the fit results using the first cumulant with the values using the
first two cumulant for the experimental and simulated data, the values of
the mean decay constant can be thought to be equal. In order to avoid the
contradiction that the values of $\mu _{2}$ are often negative, the apparent
hydrodynamic radius $R_{h,app}$ is obtained using the first cumulant.
Meanwhile, from the analysis of cumulant, the apparent hydrodynamic radius
is obtained from the average of the term $\exp \left( -q^{2}D\tau \right) $
in distribution $G\left( R_{s}\right) $ with the weight $R_{s}^{6}P\left(
q,R_{s}\right) $. In order to explore the effects of the distribution, the
simulated data were produced as the above simulated data with the same mean
static radius 260 nm and the different standard deviations 13, 39 and 52 nm
respectively. The constant $a$ is still chosen 1.2. From this assumption,
the mean hydrodynamic radius is 312 nm. The fit results for different
standard deviations are listed in Table 4.

\begin{center}
$\underset{
\begin{array}{c}
\text{Table 4 The apparent hydrodynamic radii }R_{h,app}\text{\ of the
simulated data produced} \\ 
\text{ using the same mean static radius and different standard deviations.}
\end{array}
}{
\begin{tabular}{|c|c|}
\hline
$\sigma /\left\langle R_{s}\right\rangle $ & $R_{h,app}\left( nm\right) $ \\ 
\hline
5\% & 315.7$\pm $0.9 \\ \hline
10\% & 325.$\pm $2. \\ \hline
15\% & 339.4$\pm $0.9 \\ \hline
20\% & 356.$\pm $1. \\ \hline
\end{tabular}
\ }$
\end{center}

From the results of apparent hydrodynamic radius, the values are obviously
influenced by the values of standard deviation. As shown in Eq. \ref{Grhrs},
the quantity $\exp \left( -q^{2}D\tau \right) $ is determined by the
hydrodynamic characteristics of particles while $R_{s}^{6}P\left(
q,R_{s}\right) $ is determined by the optical features of particles. As a
result, $g^{\left( 2\right) }\left( \tau \right) $ is determined by both the
optical and hydrodynamic characteristics of particles. When the cumulant
method is used, the apparent hydrodynamic radius $R_{h,app}$ obtained from
the normalized time auto-correlation function of the scattered light
intensity $g^{\left( 2\right) }\left( \tau \right) $ is a composite size. If
the simple size information need to be obtained from $g^{\left( 2\right)
}\left( \tau \right) $, the relationship between the optical and
hydrodynamic quantities of particles must be considered. The accurate
relationship between the static and hydrodynamic radii can be further
explored.

From above discussion, three different particle sizes can be obtained from
the light scattering techniques. The static radius is determined by the
optical characteristics, the hydrodynamic radius is obtained from the
hydrodynamic features and the apparent hydrodynamic radius is determined by
both the optical and hydrodynamic characteristics of particles. The function
between the SLS and DLS can be built if the relationship between the optical
and hydrodynamic quantities of particles can be understood.

Fig. 1 The expected and experimental values of the normalized time
auto-correlation function of the scattered light intensity $g^{\left(
2\right) }\left( \tau \right) $. Figures 1.a and 1.b show the results of the
polystrene latex and PNIPAM\ samples respectively. The symbols show the
experimental data and the line shows the calculated values with the simple
assumption $R_{h}=aR_{s}$.

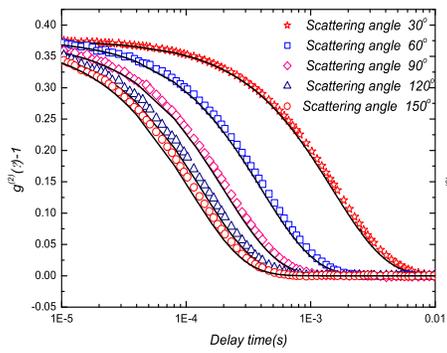
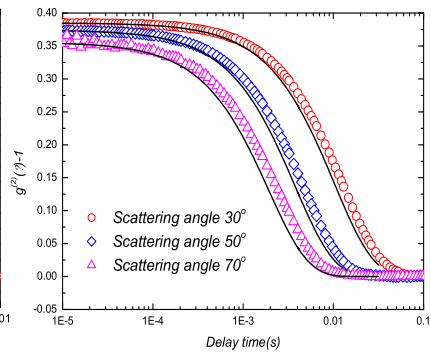
Fig. 2. The expected and experimental values of the normalized time
auto-correlation function of the scattered light intensity $g^{\left(
2\right) }\left( \tau \right) $. Figures 2.a and 2.b show the results of the
polystrene latex and PNIPAM\ samples respectively. The symbols show the
experimental data and the line shows the calculated values with the simple
assumption $R_{h}=R_{s}$.

$\left[ 1\right] $ P. N. Pusey in Neutrons, X-rays and Light: Scattering
Methods Applied to Soft Condensed Matter, edited by P. Lindner and Th. Zemb,
Elsevier Science B.V., Amsterdam, The Netherlands, 2002.

$\left[ 2\right] $ Y. Sun, Unpublished (please see my second paper)

$[3]$ C. B. Bargeron, J. Chem. Phys. 1974, 61, 2134.

$\left[ 4\right] $ B. J. Berne and R. Pecora, Dynamic Light Scattering,
Robert E. Krieger Publishing Company, Malabar, Florida, 1990.

$\left[ 5\right] $ J. C. Brown, P. N. Pusey, R. Dietz, J. Chem. Phys. 1975,
62, 1136.

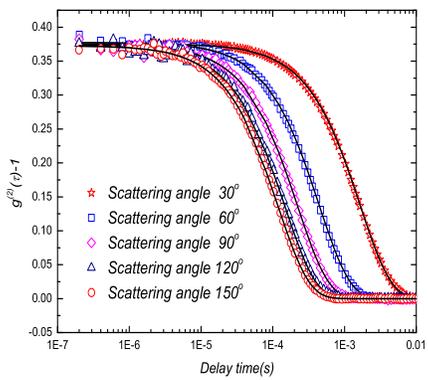
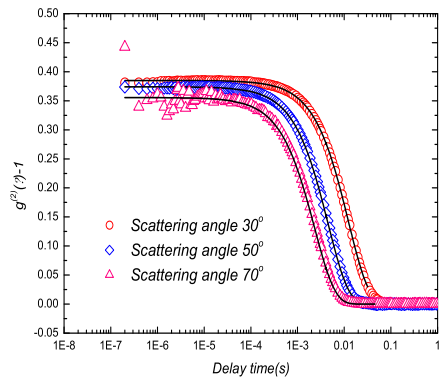
\end{document}